\documentclass[aps,prb,showpacs,twocolumn,floatfix]{revtex4}
\usepackage{graphicx,color,hyperref}

\begin{document}
\title{Signatures of critical full counting statistics in a quantum-dot chain}
\author{Torsten Karzig}
\author{Felix von Oppen}
\affiliation{Dahlem Center for Complex Quantum Systems and Fachbereich Physik, Freie Universit\"at Berlin, Arnimallee 14, 14195 Berlin, Germany}
\date{\today}
\begin{abstract}
We consider current shot noise and the full counting statistics in a chain of quantum dots which exhibits a continuous non-equilibrium phase transition as a function of the tunnel couplings of the chain with the electrodes. Using a combination of analytical and numerical methods, we establish that the full counting statistics is conventional away from the phase transition, but becomes, in a well-defined sense, essentially non-Gaussian on the critical line, where the current fluctuations are controlled by the dynamic critical exponent $z$. We find that signatures of the critical full counting statistics persist in quantum-dot chains of finite length. 
\end{abstract}
\maketitle

\section{Introduction}

It is by now well established that non-equilibrium current fluctuations in nanoscopic conductors yield much information which is not contained in the current-voltage characteristics. Most prominently, the second moment of the non-equilibrium current fluctuations, known as shot noise,\cite{Blanter} is sensitive to effects of quantum statistics on the current flow and provides access to the charge of excitations. Experiments accessing the higher moments of the current fluctuations are now becoming available,\cite{Reulet} in some cases including measurements of the entire full counting statistics.\cite{Ensslin,Haug} 

The full counting statistics \cite{Levitov, Nazarov} (FCS) generalizes the concept of photon counting statistics in quantum optics to nanoscopic conductors and characterizes the current fluctuations by means of the entire distribution function $P_T(Q)$ of the charge $Q$ passing through the conductor during a time interval of length $T$. The FCS has been investigated in a wide variety of systems, including but not limited to superconducting \cite{Belzig} and normal-superconductor hybrid systems,\cite{Muzykantskii} tunnel junctions,\cite{Shelankov} chaotic cavities,\cite{Jong} spin-correlated systems,\cite{Lorenzo} quantum dots in the Coulomb blockade \cite{Bagrets} and Kondo regimes,\cite{Gogolin} single-molecule junctions,\cite{Koch, Imura} and nanoelectromechanical systems.\cite{Pistolesi}

A common feature of the FCS in all of these systems is that any cumulant $\langle\langle Q^n\rangle\rangle$ of the FCS is proportional to $T$. This implies that in a well-defined sense, the FCS is essentially a Gaussian distribution with only small deviations. In fact, if we express the FCS in terms of the variable $q=(Q - \langle Q \rangle)/\sqrt{T}$ which measures the fluctuations of the transferred charge in units of its typical magnitude as determined by the variance $\sqrt{\langle\langle Q^2\rangle\rangle}\sim \sqrt{T}$, then all cumulants $\langle\langle q^n\rangle\rangle$ of the rescaled FCS $P_T(q)$ other than the variance tend to zero as $T\to \infty$. 

It is the purpose of the present paper to show that the FCS of a chain of quantum dots can behave in a fundamentally different manner, having cumulants ($n\geq2$) which scale as 
\begin{eqnarray}
  \langle\langle Q^n\rangle\rangle \sim T^{n/3}
\label{moments}
\end{eqnarray}
in an appropriate ``thermodynamic limit." This means that {\it all} cumulants of the FCS, when rescaled by the variance as described above, are of the {\it same} order, implying that the FCS is essentially non-Gaussian. This result is a consequence of a continuous non-equilibrium phase transition occurring in the system. Its critical indices are known exactly and the exponent appearing in Eq.\ (\ref{moments}) can be identified with $n/2z$ where $z=3/2$ is the dynamic critical exponent. We explore the FCS of the system and its consequences for current noise by a combination of analytical and numerical techniques, paying particular attention to an analysis of finite-size corrections which would be relevant in experimental realizations. 

\begin{figure}
\includegraphics[width=0.9\columnwidth]{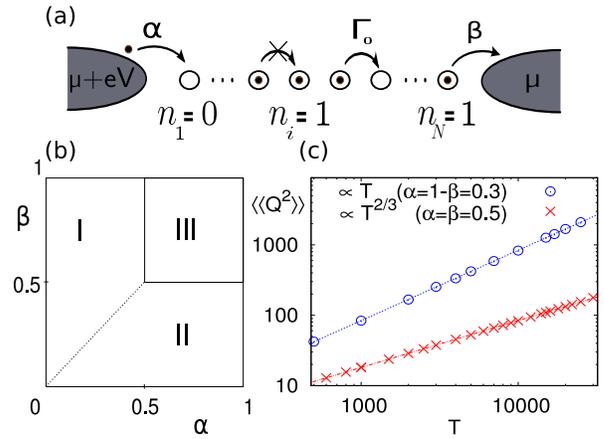}
\caption{(Color online) (a) Schematic depiction of the quantum-dot chain. (b) Mean-field phase diagram of the dot occupations.
(c) $T$-dependence of variance $\langle\langle Q^2\rangle\rangle$ both on ($\alpha=\beta=0.5$) and away from ($\alpha=0.3$, $\beta=0.7$) the critical line.
\label{QDs}}
\end{figure}

\section{Model}

Consider a series of $N$ weakly coupled metallic quantum dots as shown in Fig.\ \ref{QDs}, under the following conditions: (i) The voltage between neighboring dots is sufficiently large and energy relaxation within the dots fast enough that transport is unidirectional, i.e.\ electrons always tunnel in the direction of the voltage bias (say to the right). (ii) Due to the Coulomb blockade, the quantum dots can only switch between two neighboring charge states which we denote by $n=0$ and $n=1$. (iii) The tunneling rates between dots are equal to $\Gamma_0$, while the rate for entering the first dot from the left reservoir (leaving the last dot to the right reservoir) is tunable (e.g.\ by a gate electrode) and equal to $\alpha$ ($\beta$). A more detailed discussion of these conditions including the electrostatics of the quantum-dot chain is relegated to the Appendix. When ignoring correlations between the occupations $n_i$ of different quantum dots within a mean-field description, the rate of tunneling between quantum dots $i$ and $i+1$ is given by $\Gamma_0 n_i(1-n_{i+1})$. Thus, we can describe transport through this quantum-dot chain by the rate equations
\begin{eqnarray}
  \frac{{\rm d}n_i}{{\rm d}t}= \Gamma_0 \left[n_{i-1}(1-n_i)-n_i(1-n_{i+1})\right]+\delta J_i-\delta J_{i+1}
\label{MFT}
\end{eqnarray}
for $2\leq i \leq (N-1)$. The occupations of the first and last dot are similarly determined by the equations $\text{d}n_1/\text{d}t=\alpha(1-n_1)-\Gamma_0 n_1(1-n_2)+\delta J_1-\delta J_2$ and $\text{d}n_N/\text{d}t=\Gamma_0 n_{N-1}(1-n_N)-\beta n_N+\delta J_N-\delta J_{N+1}$. Since we will be interested in computing current fluctuations, we have already included Langevin sources $\delta J_i$ into the rate equations which account for the stochastic nature of the tunneling processes. In analogy with the Boltzmann-Langevin approach to fluctuations of the distribution function,\cite{Kogan, Kogan_book} the Poisson nature of the tunneling processes gives $\langle \delta J_i(t)\delta J_j(t')\rangle = \Gamma_0 n_{i-1}(1-n_i)\delta_{ij}\delta(t-t')$ for the correlation function of the Langevin sources. 

The stationary states of Eq.\ (\ref{MFT}) are well known from studies \cite{Derrida_92,Schutz} of the totally asymmetric simple exclusion process (TASEP) of which the quantum-dot chain under consideration is a particular realization. Remarkably, the stationary states exhibit different phases as function of $\alpha$ and $\beta$. This is  summarized in the phase diagram Fig.\ \ref{QDs} which exhibits low- and high-density phases I and II, which are separated from a maximal current phase III by a {\it continuous} non-equilibrium phase transition. (The transition between the high- and low-density phases is first order.) These results are well established by exact solutions and simulations,\cite{Derrida_92,Schutz} but can also be obtained in essence at mean-field level.

The mean-field occupation profile $n_i$ can be derived from a recursion relation based on current conservation $c=n_i (1-n_{i+1})$.\cite{Derrida_92} (Here and in the following, we use units of time such that $\Gamma_0=1$.) One finds that the occupations $n_i = \alpha$ ($n_i= 1-\beta$) in the low-density (high-density) phase are constant in the bulk. Near the boundary of the chain, they relax to the bulk value (denoted by ${\bar n}$ in the following) within a distance $\xi=1/2(1-2{\bar n})$. The divergence of the correlation length $\xi$ at $\bar n=1/2$, i.e., along the phase transition line between the high- or low-density phase and the maximal-current phase, signals the occurrence of a continuous phase transition. In analogy with second-order phase transitions in thermal equilibrium, $\xi$ can also be extracted from a linearized mean-field description of the density-density fluctuations of the model based on Eq.\ (\ref{MFT}).\cite{Pierobon} From these results, one also concludes that the respective average currents through the chain are equal to $c=\alpha(1-\alpha)$ ($c=\beta(1-\beta)$).

The current fluctuations, including the entire distribution function, have been discussed for several related models in a number of publications.\cite{Praehofer,Sasamoto,Derrida98,Bodineau04,Bertini05,Bodineau05} These works focused on exact solutions for infinite chains and ring geometries with particular choices of initial conditions. Here, we employ an alternative field theory description starting from (the approximate) Eq.\ (\ref{MFT}) which builds on existing approaches to the FCS of nanoscopic conductors \cite{Gutman} and which also provides a convenient starting point to discuss finite-size systems.

\section{Field-theory description of the FCS in the thermodynamic limit}

In terms of the charge $Q_i$ passing the bond between quantum dots $i-1$ and $i$,
\begin{eqnarray}
  Q_i = \int_0^T {\rm d}t [n_{i-1}(1-n_i)+\delta J_i],
\label{chargeT}
\end{eqnarray}
the FCS (generalized to the joint distribution function of all the $Q_i$) is defined by 
\begin{eqnarray}
   P_T(\{Q_i\})\!\! &=&\!\! \left\langle \!\prod_i \delta\!\left(\!Q_i \!-\!\! \int_0^T \!\!\!{\rm d}t [n_{i-1}(1-n_i)+\delta J_i]\!\right)\!\!\right\rangle_{\delta J}.
   \label{FCS_start}
\end{eqnarray}
In performing the average over the $\delta J$, we need to remember that the occupations $n_i$ must satisfy the Boltzmann-Langevin equation (\ref{MFT}) and are thus themselves dependent on the $\delta J$. We can formally average independently over occupations and Langevin sources after enforcing the Boltzmann-Langevin equation by means of $\delta$-functionals in the average. Doing so, passing to the Fourier transform ${\tilde P}_T(\{\chi_i\}) = \int \prod_i {\rm d}Q_i \exp({\rm i}\sum_j \chi_j Q_j) P_T(\{Q_j\})$, and exponentiating the functional $\delta$-functions, we obtain \cite{Gutman}
\begin{eqnarray}
{\tilde P}_T(\{\chi_i\}) = \frac{1}{\cal Z} \int \prod_i [{\rm d}\alpha_i(t)]\left\langle {\rm e}^{{\rm i}S}\right\rangle_{n,\delta J}
\end{eqnarray}
with the action 
\begin{widetext}
\begin{equation}
  S = \sum_j \int \text{d}t \left\{ \chi_j [n_{j-1}(1-n_j)+\delta J_j]+\alpha_j\left[\frac{\partial n_j}{\partial t} - n_{j-1}(1-n_j) + n_j (1-n_{j+1}) - \delta J_j + \delta     
  J_{j+1}\right]\right\}.
\end{equation}
The prefactor ${\cal Z}$ ensures the normalization condition ${\tilde P}_T(\{\chi_i=0\})=1$. Moreover, we have made the counting fields $\chi_j$ time dependent for simplicity of notation. ${\tilde P}_T(\{\chi_j\})$ follows by setting $\chi_j(t) = \chi_j$ for $0\leq t\leq T$ and zero otherwise. Performing the Gaussian average over $\delta J$, passing to the continuum limit (with the distance between quantum dots set to $a=1$), and integrating over the constraint field $\alpha_j$, we obtain after lengthy, but straight-forward manipulations
\begin{equation}
  S = - \int {\rm d}x {\rm d}t \left\{ \frac{}{}\chi \partial_x^{-1}\partial_t n+\frac{{\rm i}}{2c} \left(\partial_t n + \partial_x j -\frac{1}{2}\partial_x^2 n\right){\partial^{-2}_x}\left(\partial_t n + \partial_x j -\frac{1}{2}\partial_x^2 n\right)\right\}
\label{action}
\end{equation}
\end{widetext}
where $c=\bar n(1-\bar n)$ denotes the average current in terms of the average occupation $\bar n$. For simplicity, we removed a trivial term from this action such that its Fourier transform gives a shifted FCS defined as the distribution function of $Q-cT$, a quantity with zero average. In the following, this shifted charge variable will be denoted by $Q$.

It is important to note that the action Eq.\ (\ref{action}) is {\it non-Gaussian} in the occupation field $n$ due to the presence of the terms involving $j=n(1-n)$. Writing $n=\bar n+\Delta n$, we have $\partial_x j = (1-2\bar n)\partial_x\Delta n - 2 \Delta n\partial_x \Delta n$. One finds by power counting that the non-linear convective terms are irrelevant away from the critical line well inside the high- and low-density phases where $\bar n \neq 1/2$, but become relevant at the continuous phase transition where $\bar n=1/2$. We now turn to an analysis of the FCS  based on the action Eq.\ (\ref{action}).

Employing the action (\ref{action}) in the absence of the non-linear convective terms yields a Gaussian FCS which is fully described by the variance of $Q$. Using $\langle\langle Q^2\rangle\rangle = -\frac{\partial}{\partial \chi^2}\ln {\tilde P}_T(\chi)|_{\chi=0}$, we obtain 
\begin{eqnarray}
  \langle\langle Q^2 \rangle\rangle = c\int_{-T/2}^{T/2} {\rm d}t {\rm d}t' \int\frac{d\omega}{2\pi}\frac{1}{N}\sum_q \frac{\omega^2{\rm e}^{-{\rm i}\omega(t-t')}}{(\omega - vq)^2+q^4/4}
\end{eqnarray}
in terms of the drift velocity $v=1-2\bar n$. Performing the time and frequency integrations gives
\begin{eqnarray}
  \langle\langle Q^2 \rangle\rangle = 2c \int\frac{\text{d}q}{2\pi} \frac{1-{\rm e}^{-q^2T/2}\cos(vqT)}{q^2}.
\label{variance}
\end{eqnarray}
For nonzero $v$, i.e., well inside the high- and low-density phase, the exponential factor in the numerator can be replaced by unity and we obtain $\langle\langle Q^2\rangle\rangle = c v T$. 

The result for $\langle\langle Q^2\rangle\rangle$ is strikingly different on the critical line where $v=0$. Here, one finds from Eq.\ (\ref{variance}) (i.e., within linearized mean-field theory) that $\langle\langle Q^2\rangle\rangle = c \sqrt{2T/\pi}$. The crucial observation is that the variance no longer scales proportional to $T$. Eq.\ (\ref{variance}) suggests that we can identify the exponent of $T$ in this relation as $1/z$ in terms of the dynamic critical exponent $z$ which takes the value $z=2$ within linearized mean-field theory. 

Indeed, this identification allows us to extend our results beyond linearized mean-field theory. For $v=0$ the full action Eq.\ (\ref{action}) is closely related to the one-dimensional Burgers equation 
\begin{eqnarray}
  \partial_t \phi - 2\phi \partial_x \phi -\frac{1}{2}\partial_x^2\phi = \partial_x \eta 
\end{eqnarray}
for a field $\phi(x,t)$ driven by a Langevin source $\eta(x,t)$ with correlator $\langle\eta(x,t)\eta(x',t')\rangle \sim \delta(x-x')\delta(t-t')$. This suggests that the true variance of the quantum-dot chain obeys $\langle\langle Q^2\rangle\rangle \sim T^{1/z}$, where $z$ is the exact dynamic critical exponent $z=3/2$ of the Burgers equation.\cite{Forster} This is confirmed by our numerical simulations as shown in Fig.\ \ref{QDs}c.

\begin{figure}
	\includegraphics[angle=270,width=1\columnwidth]{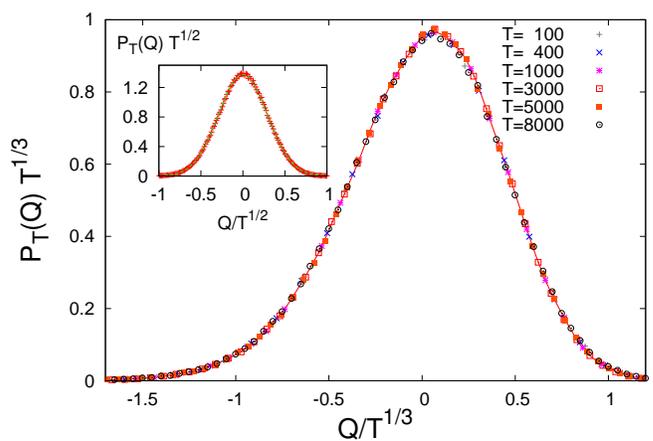}
\caption{(Color online) Full counting statistics of the totally asymmetric exclusion process vs a scaled charge variable $Q$ on the critical line (main panel) [$\alpha=\beta=0.5$], where the FCS is essentially non-Gaussian and far from the phase transition (inset) [$\alpha=0.3$, $\beta=0.7$], where the scaled FCS converges to a Gaussian for large $T$.\label{FCS}}
\end{figure}

Within linearized mean-field theory, the FCS satisfies the scaling relation $P_T(Q) = T^{-1/4} f(Q/T^{1/4})$ in terms of a scaling function $f(x)$. This can be seen directly from the linearized version of the action Eq.\ (\ref{action}) with $v=0$ by noting that it remains invariant under the rescalings $t\to t/T$, $x\to x/T^{1/2}$, $n\to nT^{1/4}$, and  $\chi\to \chi T^{1/4}$. In view of our results for the variance, one may then expect that the {\it exact} FCS satisfies the scaling relation
\begin{equation}
  P_T(Q) = \frac{1}{T^{1/2z}}\, f_\infty\!\left(\frac{Q}{T^{{1}/{2z}}}\right)
\label{FCS_scaling}
\end{equation}
with $z=3/2$. This is indeed nicely confirmed by our numerical simulations as shown in Fig.\ \ref{FCS}. This scaling relation is at the heart of the anomalous FCS of the quantum-dot chain under considerations. Indeed, the result for the cumulants in Eq.\ (\ref{moments}) is an immediate consequence of this relation. 

\begin{figure}
\includegraphics[angle=270,width=1\columnwidth]{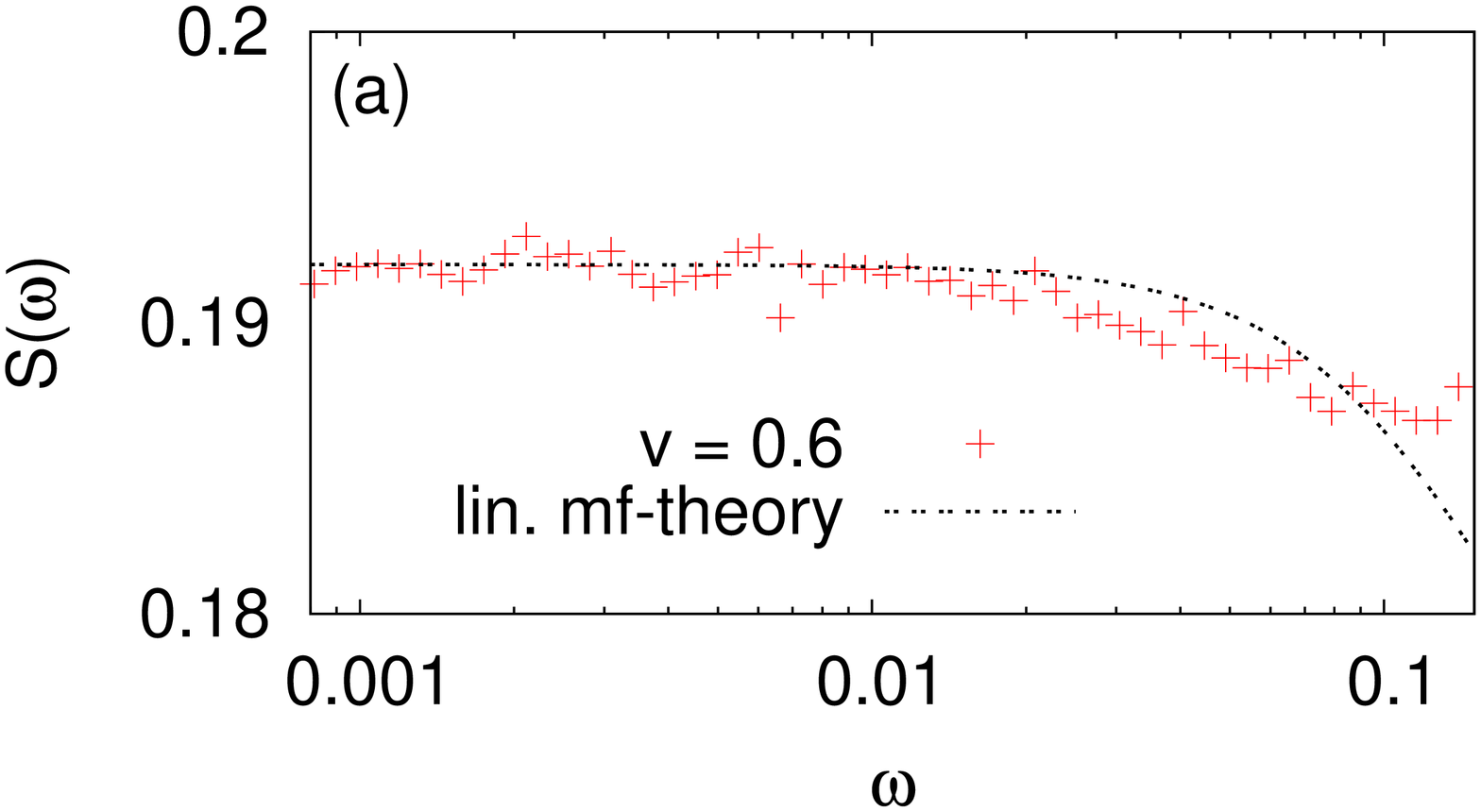}\\
\includegraphics[angle=270,width=1\columnwidth]{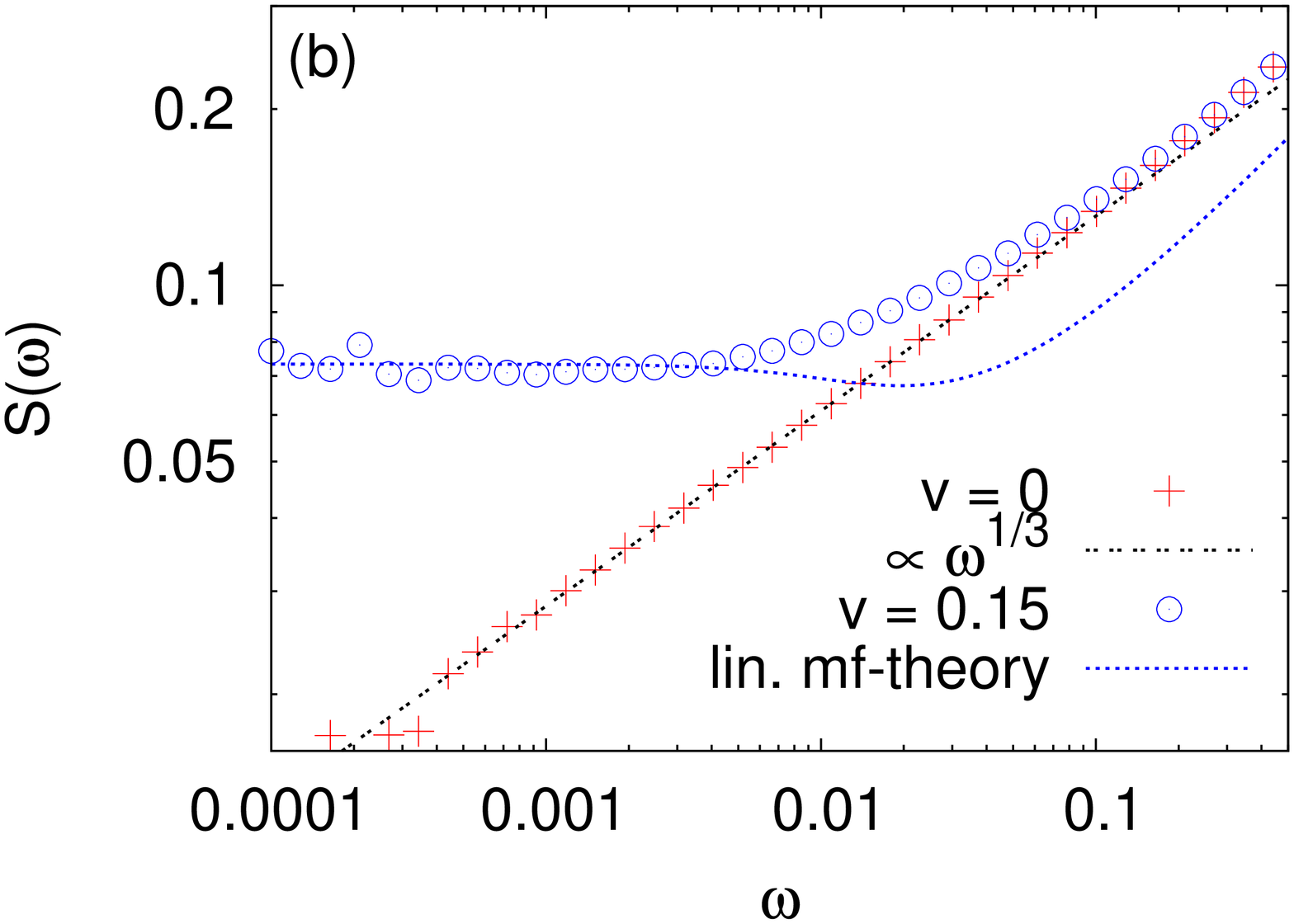}
\caption{(Color online) Current noise as function of frequency $\omega$ (a) well away from the critical line (with parameters $\alpha = 0.2$,\ $\beta=1-\alpha$) and (b) at (near) the critical line with parameters $\alpha = 0.5$ ($\alpha=0.425$) and $\beta=1-\alpha$. The simulation data are compared with the predictions of linearized mean-field theory (Eq.\ \ref{Slin}) and the critical scaling $S(\omega)\propto\omega^{1/3}$.\label{Noise} The assignment of symbols and lines is given in the legends within the figures.}
\end{figure}

From the point of view of current experiments, it is also interesting to calculate the power spectrum $S(\omega) = 2 \int {\rm d}t {\rm e}^{{\rm i}\omega \tau} \langle \delta I(t)\delta I(t+\tau)\rangle$ of the current $I(t)$. Far from the critical line, we readily obtain 
\begin{equation}
  S(\omega) = 2c \frac{1}{N}\sum_q \frac{\omega^2}{(\omega - vq)^2+q^4/4}
\end{equation}
within linearized mean-field theory. Evaluating the sum over $q$ yields
\begin{equation}
S(\omega)= c[v^{2}{\rm Re} (v^{2}-2{\rm i}\omega)^{-1/2}+{\rm Re} (v^{2}-2{\rm i}\omega)^{1/2}] 
\label{Slin}
\end{equation}
which leads to a constant noise power $S(\omega)=2 c v$ for small $\omega$. Sufficiently far from the critical line, this result is in good agreement with our simulations, see Fig.\ \ref{Noise}(a). 

On the critical line, linearized mean-field theory predicts
\begin{equation}
S(\omega)=c\omega^{1/2}.
\end{equation}
The particular value of the exponent in this relation is an artifact of linearized mean field theory. The true scaling of $S(\omega)$ at small frequencies can be extracted from the relation $\langle\langle Q^2\rangle\rangle = \int ({\rm d}\omega/\pi \omega^2) S(\omega) \sin^2(\omega T/2)$ combined with Eq.\ (\ref{moments}) from which we read off that $S(\omega) \propto \omega^{1-1/z}$. This is confirmed nicely by our simulations, as shown in Fig.\ \ref{Noise}(b).  

In the vicinity of the critical line, i.e., for a small but finite drift velocity $v$, the noise power $S(\omega)$ crosses over between the linearized mean-field theory results for small frequencies and the critical behavior for larger frequencies, see Fig.\ \ref{Noise}(b). This can be understood in terms of the correlation length $\xi=1/2v$. Once $\omega > v^2$, excitations do not drift far enough during a time $\omega^{-1}$ to explore the finite correlation length of the system, $v/\omega < \xi$. As a result, linearized mean-field behavior applies only for $\omega< v^2$ while the system appears critical for $\omega > v^2$. 

\section{Finite-size corrections}

Any experimental realization would contain only a finite number $N$ of quantum dots. In this section, we discuss to which degree signatures of the critical full counting statistics persist in finite-size systems. For finite $N$, one expects that the full counting statistics obeys Eq.\ (\ref{FCS_scaling}) up to times $T \sim N^z$. Within linearized mean-field theory, this can be obtained by the following argument. In a finite-length system, the upper limit for the correlation length $\xi=1/2v$ is $\xi=N$, from which we can extract an effective drift velocity $v$ given by $v_{\rm eff}=1/(2N)$. Thus, we expect finite-size effects to become relevant once $v_{\rm eff}T=N$ which immediately yields the above estimate with $z=2$. Thus, the ano\-malous scaling persists up to values of $T$ which are large multiples of the microscopic tunneling rate  $1/\Gamma_0$, even for moderate values of $N$.

\begin{figure}
\includegraphics[angle=270,width=1\columnwidth]{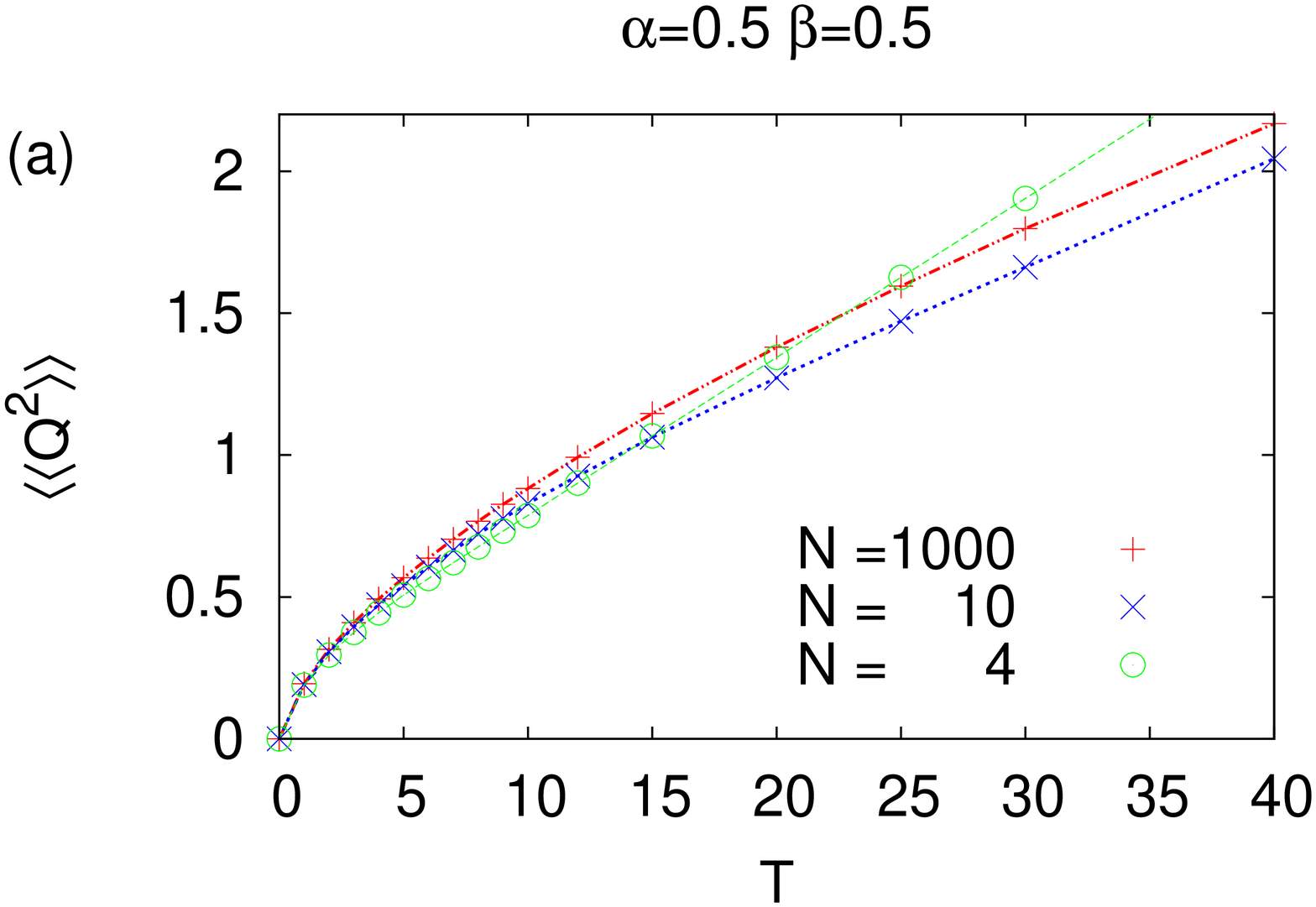}
\includegraphics[angle=270,width=1\columnwidth]{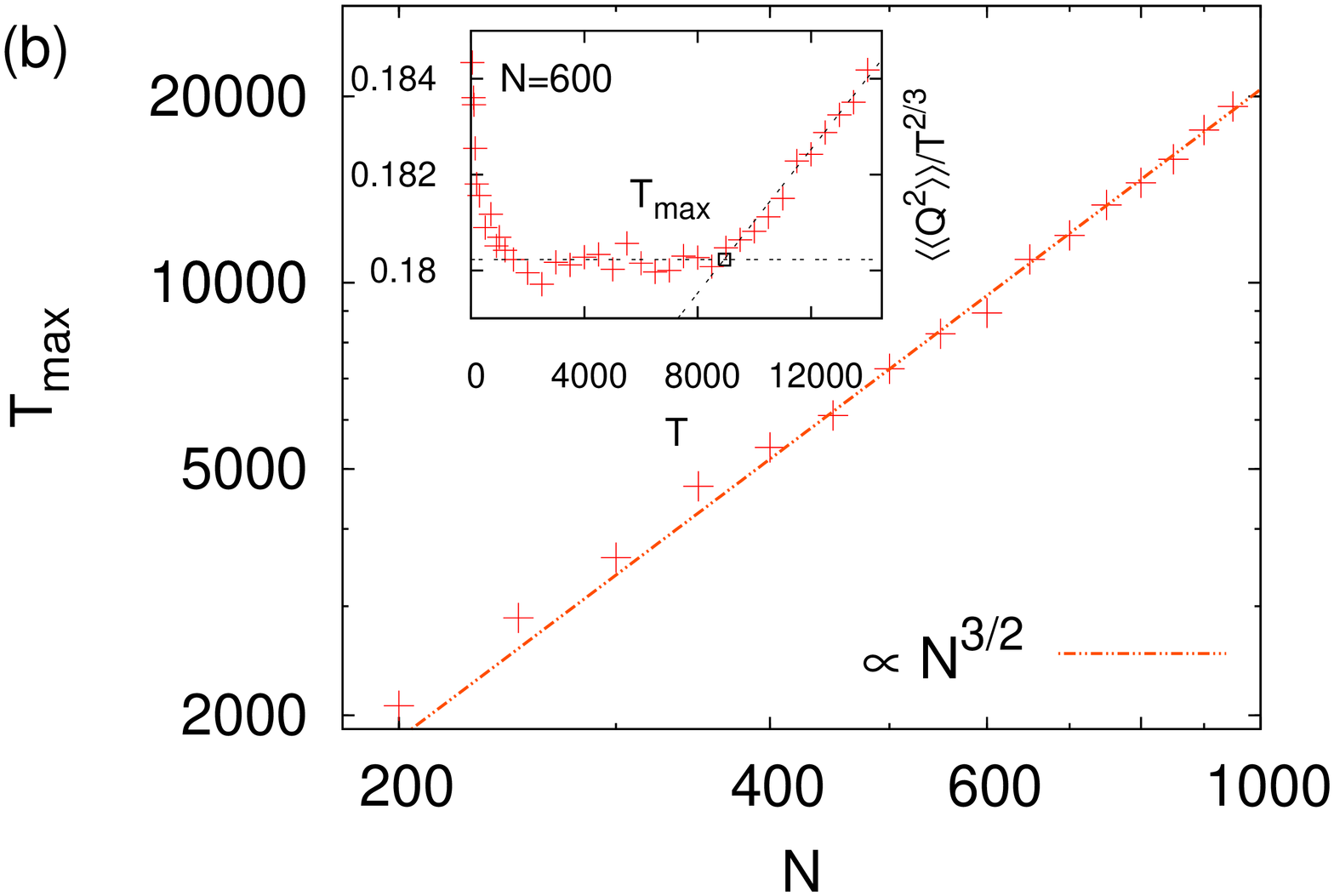} 
\caption{(Color online) (a) Dependence of variance $\langle\langle Q^{2} \rangle \rangle$  on time $T$ for small system sizes $N=4$ and $10$. For comparison, a system with $N=1000$ is also shown. (b) Crossover time $T_{\rm max}$ up to which critical scaling persists as function of system size $N$. The points are extracted by plotting $\langle\langle Q^{2}\rangle\rangle/T^{2/3}$ vs $T$ at each $N$ and identifying the point where the curve deviates from a constant, as illustrated in the inset for a particular value of $N$. (Note that the linear fit of the curves for $T>T_{\rm max}$ merely serves to identify $T_{\rm max}$ and does not reflect a theoretically expected result.)} 
\label{Nfinite}
\end{figure}

It turns out that the signatures of the critical electron dynamics remain most pronounced in the variance and the noise power. This is shown in Fig.\ \ref{Nfinite}(a) which exhibits the variance $\langle\langle Q^2\rangle\rangle$ as function of $T$ for small system sizes. It is apparent from the plots that a sublinear $T$ dependence is observed for small $T$, for chains as short as $N=4$. The dependence is close to the non-linear behavior expected in the infinite system when $T$ is not too large, and crosses over to linear behavior only for larger $T$ in accordance with the discussion in the previous pargraph. Indeed, we can extract the maximal $T$ up to which critical behavior persists from $\langle\langle Q^2\rangle\rangle$ by plotting $\langle\langle Q^2\rangle\rangle/T^{2/3}$ vs $T$, see the inset of Fig.\ \ref{Nfinite}(b). In this plot, critical full counting statistics is indicated by a constant while deviations from the constant at large $T$ indicate significant finite-size corrections. Extracting a $T_{\rm max}$ for different $N$ and plotting $T_{\rm max}$ vs $N$ nicely fits the expected $T_{\rm max}\sim N^{3/2}$ dependence with the exact exponent $z=3/2$, see Fig.\ \ref{Nfinite}(b).

\begin{figure}
\includegraphics[angle=270,width=1\columnwidth]{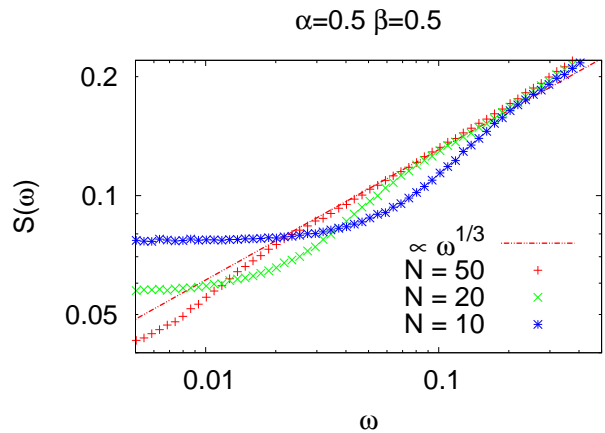}
\caption{(Color online) Noise power as function of frequency for quantum-dot chains with $N=10,20$ and $50$. Note that one recovers the critical scaling for sufficiently large frequency for all system sizes. \label{FiniteSize}}
\end{figure}

Experimentally, it may be more accessible to study the noise power as a function of frequency. At small frequencies
$\omega<T_{\rm max}^{-1}\sim1/N^z$, the behavior should be noncritical while the critical scaling $S(\omega)\sim \omega^{1/3}$ would persist at larger frequencies. This expectation is borne out by our numerical results, see Fig.\ \ref{FiniteSize}.

\begin{figure}
\begin{centering}
\includegraphics[angle=270,width=1.05\columnwidth]{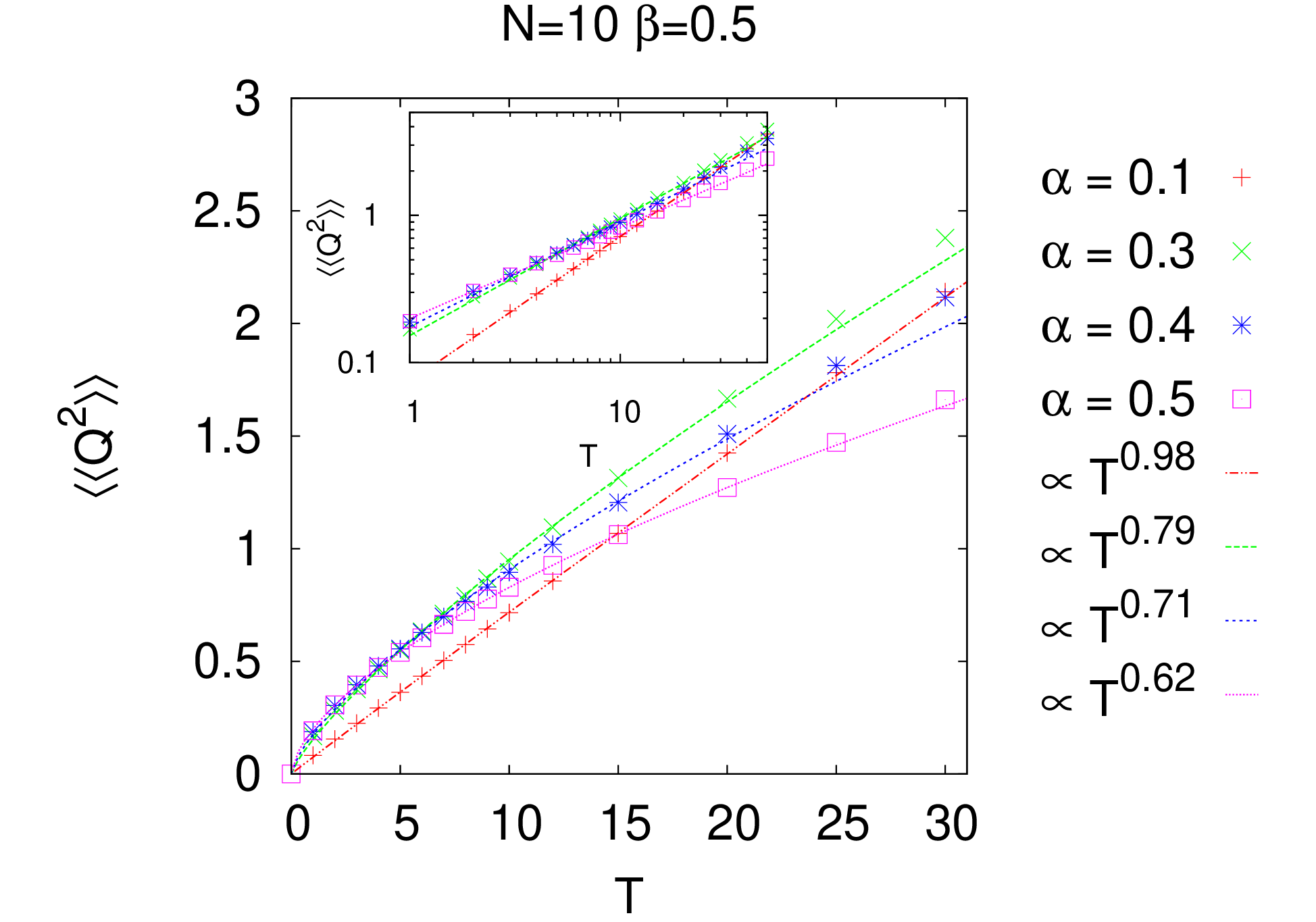}
\par\end{centering}
\caption{(Color online) Variance $\langle\langle Q^{2} \rangle\rangle$ as function of $T$ for finite $N=10$, at various distances from the nominal critical line. Sublinear behavior is observed in the entire critical region, as explained in the text. Within the critical region, the dependence be approximately fitted to a power law, with exponents close to 2/3, as illustrated in the inset. 
\label{fig:alpha-dependence}}
\end{figure}

One also expects the transition to smear over a certain range of tunneling rates $\alpha$ and $\beta$ when $N$ is finite, implying that in finite-size systems signatures of the critical electron dynamics should be visible even away from the nominal transition line. The region over which the transition becomes smeared can be estimated as follows. As argued above, 
critical behavior is visible when $vT_{\mathrm{max}}\lesssim N$, i.e., for $v\lesssim N^{1-z}$. Using that $v=1-2\bar{n}$, we find a critical region $|\bar{n}-1/2|\lesssim1/(2N^{1/2})$. This estimate is consistent with our numerical results as shown in Fig.\ \ref{fig:alpha-dependence}. This figure shows that the dependence of the variance $\langle\langle Q^{2} \rangle\rangle$ on $T$ exhibits sublinear behavior in the entire critical region. Moreover, it is demonstrated in the inset that the dependences are rough power laws with exponents close to $2/3$ throughout the critical region, while the exponent approaches unity outside of the critical region. 

\section{Conclusions}

We have shown that a chain of quantum dots may realize a totally asymmetric exclusion process and thus exhibit a non-equilibrium phase transition as function of the tunnel couplings to the leads which results in critical full counting statistics of the current fluctuations.  We find that the non-equilibrium phase transition controls the dynamics over a wide range of time scales, even in quantum-dot chains of finite length. Our minimal model assumes that the hopping rates between quantum dots are all equal. Due to the exponential sensitivity of tunneling, this would presumably be exceedingly difficult to realize in a lateral arrangement of quantum dots. Instead, it appears more promising to employ a vertical setup \cite{Tarucha} in which the tunneling barriers and the quantum dots can be formed by a well-defined number of monolayers. We expect that including weak backscattering or rare double occupation of a dot would leave our results qualitatively unchanged. Nevertheless, it would be interesting to study their consequences as well as the influence of variations in the tunneling rates in more detail.

It would also be interesting to identify other systems which exhibit a critical full counting statistics, especially systems whose dynamics is more directly controlled by quantum mechanics. An intriguing possibility is transport near a quantum phase transition. 

\begin{acknowledgments}
This work was supported in part by DIP and  Sfb 658. The authors thank the KITP (supported by the National Science Foundation under Grant No. PHY05-51164) for hospitality while part of this work was performed.
\end{acknowledgments}

\appendix

\section{Electrostatics of a quantum-dot chain\label{sec:Electrostatics}}

\begin{figure}
\begin{centering}
\includegraphics[bb=0bp 0bp 526bp 230bp,clip,width=0.9\columnwidth]{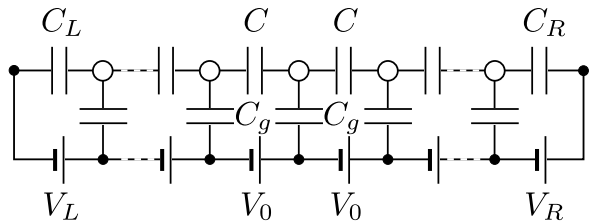}
\par\end{centering}
\caption{\label{fig:Schematic} Equivalent circuit of the quantum-dot chain. White (black) dots indicate quantum dots (electrodes).}
\end{figure}

In this appendix, we briefly discuss the electrostatics of a quantum-dot chain. We consider metallic quantum dots with a continuous spectrum so that the tunnel rates can be assumed to be a linear function of the energy gained in a tunneling process. To derive this energy gain, we model the quantum-dot chain by the equivalent circuit shown in Fig.\ \ref{fig:Schematic} which is described by the capacitance matrix $\mathbf{C}$ defined as
\begin{equation}
\mathbf{q}=\left(\begin{array}{c}
\mathbf{q_{d}}\\
\mathbf{q_{e}}\end{array}\right)=\left(\begin{array}{cc}
\mathbf{C_{dd}} & \mathbf{C_{de}}\\
\mathbf{C_{ed}} & \mathbf{C_{ee}}\end{array}\right)\left(\begin{array}{c}
\mathbf{v_{d}}\\
\mathbf{v_{e}}\end{array}\right)=\mathbf{C}\mathbf{v}
\end{equation}
where the index $\mathrm{d}$($\mathrm{e}$) enumerates the quantum dots (electrodes). The charges and the potentials on the dots and electrodes are denoted by ${\bf q}$ and ${\bf v}$, respectively 

Up to irrelevant constants, the energy of the system takes the form
\begin{equation}
U=\frac{1}{2}\mathbf{v_{d}}^{T}\mathbf{C_{dd}}\mathbf{v_{d}}.
\end{equation}
In an infinite chain, the energy difference $\Delta U_{i}$ induced by a tunneling event from the $i$th to the $(i+1)$th dot becomes
\begin{eqnarray}
\Delta U_{i} & \simeq & -V_{0}+\frac{1+n_{i+1}-n_{i}}{C_{g}}\label{eq:U}\\
 &  & +\frac{C}{C_{g}}\left(\frac{n_{i+2}-n_{i+1}+n_{i}-n_{i-1}-1}{C_{g}}-2V_{0}\right),\nonumber 
\end{eqnarray}
to first order in the weak coupling limit $C\ll C_{g}$. Here, $\{n_{i}\}$ denote the occupation numbers of the initial state and we use units with $e=1$. The weak coupling limit $C\ll C_{g}$ ensures that the contribution to the energy arising from interdot interactions is small compared to the contribution of the gate capacitances. According to Eq.\ ($\ref{eq:U}$) we conclude that double occupation and backscattering is strongly suppressed as long as $k_BT\ll V_0\ll e^2/C_g$ while the tunneling rates are approximately equal as long as $C\ll C_g$.

\end{document}